\documentclass[twocolumn,english,letterpaper,showpacs,floatfix]{revtex4}
\begin{document}
\newcommand{\be}{\begin{equation}}
\newcommand{\ee}{\end{equation}}
\newcommand{\bear}{\begin{eqnarray}}
\newcommand{\eear}{\end{eqnarray}}
\newcommand{\dpf}{\frac}
\title{ Constraints on dark energy from holography}
\author{Bin Wang}
\email{wangb@fudan.edu.cn}
\affiliation{ Department of Physics, Fudan
University, Shanghai 200433,
P. R. China }
\author{Elcio Abdalla}
\email{eabdalla@fma.if.usp.br}

\affiliation{Instituto De Fisica, Universidade De
Sao Paulo, C.P.66.318, CEP
05315-970,  Sao Paulo, Brazil }
\author{Ru-Keng Su}
\email{rksu@fudan.ac.cn} 
\affiliation{China Center of Advanced Science and
Technology, World Lab,
P.B.Box 8730, 100080, Beijing, and Department of
Physics, Fudan
University, Shanghai 200433, P. R. China }

\begin{abstract}
Using the holographic principle we constrained the
Friedmann equation,
modified by brane-cosmology inspired terms which
accomodate dark energy
contributions in the context of extra dimensions.
\end{abstract}

\maketitle

More than twenty years ago, Bekenstein first
proposed that for any isolated 
physical system of energy $E$ and size $R$, there is
an upper bound on the
entropy $S\leq S_B=2\pi RE$ \cite{1}. In spite of
criticisms about that 
derivation \cite{2}, the Bekenstein entropy bound
has received independent
support \cite{3}. Choosing $R$ to be the particle
horizon, Bekenstein gave
a prescription for a cosmological extension of the
bound \cite{4}. 

In view of the well-known example of black hole
entropy, an influential
holographic principle was put forward ten years ago,
relating the maximum
number of degrees of freedom in a volume to its
boundary surface area
\cite{5}. For systems of limited gravity, the
Bekenstein entropy bound
implies the holographic principle. The extension of
the holographic
principle to cosmological settings was first
addressed by Fischler and
Susskind (FS) \cite{6}. Subsequently, various
different modifications of
the FS version of the holographic principle have
been raised \cite{7,8}. 
In addition to the study of holography in
homogeneous cosmologies, 
attempts to generalize the holographic principle to
a generic realistic 
inhomogeneous cosmological setting were carried out
\cite{9,10}. Other 
extensions of the holographic principle to brane
cosmology have also 
been given \cite{11}. 

The holographic principle is viewed as a real
conceptual change 
in our thinking about gravity \cite{12}. It is
interesting to note that 
holography implies the probable values of the
cosmological constant in a 
large class of universes \cite{13}. In the
inhomogeneous cosmology,
holography was also realized as a useful tool to
select physically
acceptable models \cite{10}. Recently, the idea of
holography has further
been applied to the study of inflation and gives
plausible upper limits to
the number of e-foldings \cite{14}. 

Recent observations based on type IA supernova as
well as from Cosmic
Microwave Background Radiation results imply that
the expansion of our
universe is speeding up instead of slowing down
\cite{15}. This discovery
has posed a fundamental challenge to the standard
models of both particle
physics and cosmology. Whether holography can bring
us some light in
understanding the profound puzzle posed by dark
energy is a question we
want to ask. Motivated by the suggestions in
\cite{13} \cite{16}, attempts
to apply the holography to the dark energy due to
quantum corrections
to the cosmological constant were carried out in
\cite{17} \cite{18}. It
was found intriguing that using the event horizon
size as the infrared
cutoff relevant to the dark energy, the equation of
state of the dark
energy can be predicted correctly \cite{18}. 

In addition to attributing dark energy to quantum
vacuum energy, to a light 
and scalar field or a frustrated network of
topological defects, recently 
further attention has been paid to a modification of
general relativity in
order to explain the accelerated expansion 
\cite{19,20,21}. This idea is 
attractive since it has close ties to high energy
physics such as string 
theory or extra dimensions and does not necessarily
suffer from fine
tuning problems. In this paper, we will parametrize
the physical cause of
acceleration with an arbitrary additional term in
the Friedmann equation 
introduced in \cite{20}. Employing the idea of
holography, we are going 
to derive the constraints on the equation of state
of dark energy and the 
extra-dimensional contribution to the Friedmann
equation.  

Suppose that the effects of extra dimensions
manifest themselves as a
modification to the Friedmann expansion rate
equation of the form \cite{20} 
\be\label{1}  
(H/H_0)^2=(1-\Omega_M)(H/H_0)^{\alpha}+\Omega_M (1+z)^3.
\ee
The first term on the right-hand-side of (\ref{1})
is the correction to
Friedmann equation due to infinite extra dimensions
accounting for the
geometric dark energy. $H_0$ is the Hubble parameter
at present,
$\Omega_M$ is the critical energy density of matter
and $z$ is the
redshift relating to cosmic scale factor $a$ by
$z=a^{-1}-1$. 

The model under consideration decribes an ever expanding
universe. Although a further transition is not ruled
out, this is a
good simplifying assumption. A cosmological constant
with dust and
radiation also leads to ethernal expansion. 

As an example, consider a simple, single
extra-dimensional model
\cite{19}. The effective, low energy action is given by 
\be
S=\dpf{M^2_{Pl}}{r_c}\int d^4x dy\sqrt{g^{(5)}}\Re+\int
 d^4x\sqrt{g}(M^2_{Pl}R+L_{SM}), 
\ee
where $M^2_{Pl}=1/8\pi G$ is the four-dimensional
Planck scale,
 $g^{(5)}_{AB}$ is the five-dimensional metric, $y$
is the
 extra-dimension, $\Re$ is the five-dimensional
Ricci scalar, $g$ is the
 trace of the four-dimensional metric, $R$ is the
four-dimensional Ricci
 scalar and $L_{SM}$ is the Lagrangian of the fields
in the standard
 model. For the Friedmann-Robertson-Walker Ansatz 
\be
ds^2_5=f(y,H)ds^2_4-dy^2,
\ee
where $ds^2_4$ is the four-dimensional maximally
symmetric metric, and $H$
 is the four-dimensional Hubble parameter, one gets
the modified Friedmann
 equation on the brane under the form
\be
H^2\pm \dpf{H}{r_c}=\dpf{8\pi G\rho_m}{3}.
\ee
These general features persist for an arbitrary
number of dimensions,
leading to (\ref{1}).

The general form (\ref{1}) is mathematically
equivalent to a time variable
dark energy equation of state function \cite{20}
\cite{21} 
\be \label{2}  
\omega_{D}(z)=-1+\dpf{1}{3}\dpf{d\ln(H^{\alpha}/H^{\alpha}_0)}{d\ln(1+z)}
+ \frac 13 \frac{\dot\Omega_M}{1-\Omega_M}\quad .
\ee

As a matter of fact $\Omega_M=\rho/\rho_c$ is not a
constant, but it
can be neglected. Indeed, from
$\dot{\rho}=-3H(1+\omega)\rho$, we have  
\be 
\rho=\rho_0 exp\lbrack 3(1+\omega)\int^1_a d\ln
a\rbrack\quad ,
\ee
where we have taken $\rho_0$ as the matter density
at present,
$a(t_0)=1$. For $\omega=0$, the matter dominated
era, we have $\rho=\rho_0
a^{-3}$. Then $\rho_0/\rho_c=\rho a^3/\rho_c$.
Considering $\rho_c=3M_p^2
H^2$, we have $\dpf{3M_p^2 H_0^2}{3M_p^2
H^2}=\Omega_M a^3$, then $\Omega_M
=H_0^2 H^{-2} a^{-3}=H_0^2/(a \dot{a}^2)$, which
depends on time. 

However such an additional term is very small from
now untill the distant
future. Taking $a\sim t^p$, ($p=2/(1+\omega)$, for
the expansion of the
universe due to matter), the additional term is
proportional to
$1/t^{3p-1}$, which should disappear for large $t$.
Therefore we can
neglect the third term of (\ref{2}) here. For a
distant future the last 
term in (\ref{2}) will be even smaller and can be
neglected for sure
(unless we allow new physics in the future). 

At this point it is worthwhile making some comments
concerning the fact
that in the distant future there is a further
solution of the Friedmann
equation given by de Sitter space, that is, an
exponentially growing
universe, for which our arguments fail. In fact, the
arguments given in
this paper are valid for a varying equation of
state, which can change
from some value in the past, to a value not equal to
-1 today. In case it
stabilizes in the future at the value -1, then the
scale factor of the
universe may evolve in different stages: 
\begin{itemize}
\item matter dominated era , $a\sim t^{2/3}$,
$\ddot{a}<0$.

\item after dark energy starts to play effect, the
evolution of the scale
 factor will become faster, $a\sim
t^{\alpha/[3(1+\omega_D)]}$, $\ddot{a}>0$. 

\item when the dark energy dominates, neglecting the
matter, the scale
factor evolves exponentially. 

\end{itemize}

We argue that we are living the second stage, where
the dark energy plays the
role of the expansion of the universe,  the universe
expands not as fast
as that of the purely dark energy era, so we can use
the expression of
the scale factor indicated in the second item above.

If $\omega $ in the far future is not equal to minus
one, the argument
above stops in the second phase.

In any case, our paper relies on the hidden
assumption that the $\omega$
parameter varies with time. This is a matter which
is being discussed
since some time \cite{23}.

The continuity equation still holds,
$\dot{\rho}=-3H(\rho+P)$. 
The equation of state of the universe follows
immediately,
\be\label{3}
\omega_T(z)=-1+\dpf{1}{3}\dpf{d\ln(H^2/H^2_0)}{d\ln(1+z)}.
\ee
For an equation of state with a constant
$\omega$-parameter,
$\omega=P/\rho$, the energy density varies as
$(1+z)^{3(1+\omega)}$.  It
was found that during the matter dominated era,
$\omega=-1+\alpha/2$,
while during the earlier radiation-dominated epoch,
$\omega=-1+2\alpha/3$
\cite{20}.

Without dark energy, the universe expands as $a\sim
t^{\dpf{2}{3(1+\omega)}}$. For the far distant
future, we can neglect all
matter density $\rho$, thus we have from the
Friedmann equation the
exponential expansion of the universe. This
corresponds to
$\omega_D=-1$. For the present universe, the
universe has not been
expanded that fast, and we can assume $\omega_D$
seen today as a
constant. Then due to the dark energy, the expansion
of the universe
$a\sim t^{\alpha/[3(1+\omega_D)]}$.  To experience
accelerated expansion, 
$\ddot{a}>0$, which requires
$\dpf{\alpha}{3(1+\omega_D)}>1$. 

According to Bekenstein, for a system with limited
self-gravity, the total
entropy $S$ is less or equal than a multiple of the
product of the energy
and the linear size of the system. In the present
context, we can also
examine the entropy bound of the dark energy and
give the constraint of
the model parameter $\alpha$ to explain the dark
energy at present. So the
bound we obtain can be explained as the bound on the
structure of the dark
energy seen today.  For the
cosmological setting, the
Bekenstein bound applies to a region as
large as the particle
horizon [4], which is defined by the distance
covered by the light cone
emitted at the singularity $t=0$, $L_H=a(t)r_H(t)$,
where $r_H(t)$ is the
comoving size of the horizon defined by the
condition $dt/a=dr_H$. 
\be\label{8}
r_H=\int^t_0\dpf{dt'}{a(t')}=\dpf{3(1+\omega_D)}{3(1+\omega_D)-\alpha}
t^{\dpf{3(1+\omega_D)-\alpha}{3(1+\omega_D)}}\quad .
\ee
The total entropy inside the particle horizon behaves
as $S=\sigma 
L_H^3/a^3$, where $\sigma$ is the entropy density measured in the comoving space, which is constant in time. The Bekenstein entropy is
$S_B=EL_H=\rho
L_H^4$. Thus, the ratio $S/S_B$ reads 
\be \label{9}
\dpf{S}{S_B}\sim
(\dpf{3(1+\omega_D)}{3(1+\omega_D)-\alpha})^{-1}\sigma
t^{-\dpf{1+(1-\alpha)\omega_D}{(1+\omega_D)}} \quad .
\ee
In order to satisfy the Bekenstein bound, we require
that $-1<\omega_D<1
/(\alpha-1)$ for $\alpha>1$; $\omega_D>-1$ or
$\omega_D\leq -1/(1-\alpha)$
for $0<\alpha<1$ and $\omega_D\geq -1/(1-\alpha)$ or
$\omega_D<-1$ for 
$\alpha<0$. It is easy to see that for $\alpha>1$,
the constraint on the
equation of state is too loose. It is over the range
accounting for the
dark energy. 

Furthermore, the physical comoving size of the
particle horizon should be
positive, which requires
$\dpf{3(1+\omega_D)}{3(1+\omega_D)-\alpha}>0$. 
This gives more constraints on the equation of
state. For $\alpha>1$, it 
leads to $-2/3<\omega_D<1/(\alpha-1)$, which fails
to describe the dark 
energy. For $0<\alpha<1$, the range of dark energy
state has been further 
refined to $\omega_D>-1+\alpha/3$ or $\omega_D\leq
-1/(1-\alpha)$. For 
$\alpha<0$, we have $\omega_D\geq -1/(1-\alpha)$ or
$\omega_D<-1+\alpha/3$. 
However none of the above range of the equation of
state can accommodate the
accelerated expansion, $\ddot{a}>0$. 

We delve further in the problem by replacing the
particle horizon by the
future event horizon, $L_h=ar_h$, where
$r_h(t)=\int^{\infty}_t\dpf{dt'}
{a(t')}=\dpf{3(1+\omega_D)}{\alpha-3(1+\omega_D)}t^{\dpf{3(1+\omega_D)-
\alpha}{3(1+\omega_D)}}$. The ratio $S/S_B$ now becomes 
\be\label{10}
\dpf{S}{S_B}\sim \dpf{\sigma L_h^3/a^3}{\rho L_h^4}\sim
(\dpf{3(1+\omega_D)}{-3(1+\omega_D)+\alpha})^{-1} \sigma
t^{-\dpf{1+\omega_D(1-\alpha)}{1+\omega_D}}\quad .
\ee
In order to satisfy the Bekenstein bound, we require
$-1<\omega_D<1/(\alpha
-1)$ for $\alpha>1$; $\omega_D>-1$ or $\omega_D\leq
-1/(1-\alpha)$ for $0
<\alpha<1$ and $\omega_D\geq -1/(1-\alpha)$ or
$\omega_D<-1$ for $\alpha<0$. 

To keep the physical comoving size of the event
horizon to be positive,
the range of the equation of state is confined to
$-1<\omega_D<-1+\alpha/3$ for $\alpha>0$. To meet
the observational
result, we should take $\alpha\in (0,1)$, so that
$-1<\omega_D<-2/3$. For
$\alpha<0$, $-1+\alpha/3\leq\omega_D<-1$. 

Checking the condition for the speeding up expansion
of our universe, it
is found that both of the above ranges can
accommodate accelerated expansion.

In cosmology, particle horizon (or event horizon)
refers to the entire
past (or future) light-cone. Unlike particle or
event horizons, the
cosmological apparent horizon does not refer at all
to either initial
or final moment of the universe, furthermore it is
observable. If we choose the apparent horizon as the boundary, the Bekenstein bound becomes
\be
\dpf{S}{S_B}\sim \dpf{\sigma\tilde{r}_{AH}^3/a^3}{\rho \tilde{r}_{AH}^4}\sim \dpf{\alpha}{3(1+\omega_D)}\sigma t^{-[1+\omega_D(1-\alpha)]/(1+\omega_D)},
\ee
where $\tilde{r}_{AH}=1/H$ has been taken. The requirement of satisfying the Bekenstein entropy bound together with the accelerated expansion lead to the same constraint on the parameter $\alpha$ and equation of state as taking event horizon as the boundary.

In the cosmological setting, it is usually believed
that the Bekenstein
bound is looser than the holographic bound, which is
the opposite of what
we understood for isolated system. It would be of
great interest to
investigate whether the holographic entropy bound
can give tighter
constraints on the equation of state and
extra-dimensional contributions
to the dark energy. 

Directly applying the FS version of the cosmic
holographic principle by
using the particle horizon \cite{6}, we again face
the problem that in the
range for the equation of state that we obtained,
the universe cannot
experience accelerated expansion. 

Extending the cosmic holography by considering that
the entropy cannot
exceed one unit per Planckian area of it boundary's
surface characterized
by the event horizon, we have 
\be  \label{11}
\dpf{S}{A}\sim \dpf{\sigma L_h^3/a^3}{L_h^2}\sim
\sigma\dpf{r_h}{a^2}\sim
\dpf{3(1+\omega_D)}{\alpha-3(1+\omega_D)}\sigma
t^{1-\dpf{\alpha}{(1+\omega_D)}}\quad . 
\ee
Combing the requirement that the ratio $S/A$ not
incrasing with time,
positive $r_h$ and accelerated expansion
$\ddot{a}>0$, we obtain
$-1<\omega_D<-1+\alpha/3$ for $\alpha>0$. To meet
observation, $\alpha$ is
further refined to $0<\alpha<1$, then
$-1<\omega_D<-2/3$. For $\alpha<0$,
we have $-1+\alpha/3<\omega_D<-1$. 

If we
choose the apparent horizon as the boundary, the
holographic principle
should be expressed as: the entropy inside the
apparent horizon can
never exceed the apparent horizon area \cite{ray} 
\be 
\dpf{S(t)}{A}=\dpf{\sigma Vol_{AH}(t)}{A(t)}\leq 1
\ee
where $Vol_{AH}(t)=\dpf{V(r_{AH}(t))}{a^3(t)}$
denotes the comoving volume inside the apparent
horizon. Eq(13) can be rewritten as 
\be  
\dpf{S}{A}\sim \dpf{3(1+\omega_D)}{\alpha}\sigma
t^{1-\alpha/(1+\omega_D)}
\ee
The requirement that the ratio $S/A$ does not
increase with time together with the accelerated
expansion leads us to the same constraints for the
equation of state of dark energy as we got by using
the event horizon. Choosing the apparent horizon
leading to a reasonable equation of state of dark
energy is different from the discussion in
\cite{17}, a consequence of the special dark energy
model we have supposed.

We would also like to extend the discussion by using
the Hubble entropy
bound. Following the usual holographic arguments,
one then finds that the
total entropy should be less or equal than the
Bekenstein-Hawking entropy
of a Hubble size black hole times the number $n_H$
of Hubble regions in
the universe.  The entropy of a Hubble size black
hole is roughly
$HV_H/4$, where $V_H$ is the volume of a single
Hubble region. Considering
the universe with volume $V$, one obtains an upper
bound on the total
entropy $S<S_H=HV/4$ \cite{8}. If one applies the
Hubble entropy bound to
a region of size $L_H$, it is interesting to find
$S_H=(S_B
S_{FS})^{1/2}$.  

Employing the Hubble entropy bound to the region of
size $a$, the ratio
\be\label{12}
\dpf{S}{S_H}\sim \dpf{4\sigma a^3}{HV}\sim 
a/\dot{a}\sim t\quad .
\ee
Thus with the assumed domination of the
gravitational correction in the
Friedmann equation, Hubble entropy bound will be
violated. This is not
surprising, since not like the Bekenstein bound and
the FS bound, the
Hubble entropy bound is appropriate in the strong
self-gravitating
universe $Ha>1$ \cite{8}. With the extra-dimensional
contribution to the
dark energy to account for the accelerated
expansion, the universe will be
with limited self gravity.


\begin{table}
\caption{} 
%
%
\begin{tabular}{|c|c|c|}
%
%
{\footnotesize Boundary} & {\footnotesize 
Bekenstein Bound} &
{\footnotesize Holographic Bound}\\
\hline 
\multicolumn{1}{|c}{\footnotesize Particle hor.}&  
\multicolumn{2}{|c|}{\footnotesize Not satisfied}\\
\hline
\multicolumn{1}{|c}{\footnotesize  Event hor.} &
\multicolumn{2}{|c|}{$\matrix{ {\footnotesize
Satisfied},\cr 
              -1<\omega_D<-2/3 \quad ({\footnotesize
for} 0<\alpha<1); \cr
               -1+\alpha/3<\omega_D<-1 \quad
               ({\footnotesize for} \alpha<0) \cr}
$}\\ \hline
\multicolumn{1}{|c}{\footnotesize Apparent hor.}&
\multicolumn{2}{|c|}{ $\matrix{{\footnotesize
Satisfied},\cr 
                  -1<\omega_D<-2/3 \quad
                  ({\footnotesize for} 0<\alpha<1);\cr 
                  -1+\alpha/3<\omega_D<-1 \quad
                  ({\footnotesize for}
\alpha<0)\cr}$}\\ \hline
\end{tabular}
\end{table}

In summary, we have used the idea of holography to
study the constraints
on the geometric dark energy. Our results are listed
in the table. Contrary to the original understanding in
the cosmological setting, we found Bekenstein
entropy bound and
holographic entropy bound play the same role in
refining the geometric
dark energy. To account for the dark energy,
$\alpha$ cannot be bigger
than unit. For $0<\alpha<1$, the equation of state
of the
extra-dimensional contribution lies in the range
$-1<\omega_D<-2/3$, which
can be treated as dark energy. For $\alpha<0$, the
extra-dimensional
effect acts as an effective phantom energy, where
$-1+\alpha/3<\omega_D<-1$. The failure by using the
particle horizon to
explain the accelerated expansion of our universe
due to geometric
contribution to the dark energy has got independent
support in \cite{18}, 
where the dark energy was attributed to the
cosmological constant. The
reason for such a failure is that the particle
horizon is related to the
early universe, when dark energy played no role.

After our work was finished, we noticed that
observational constraints
on a modified Friedmann 
equation which mimics the dark energy was studied in
\cite{oystein}. Our
holographic constraints on $\alpha$ got independent
support from their
Supernovae Type IA and CMB study.  $\alpha=0$, which
is excluded in our
holographic investigation, is also disfavoured in
the studied parameter
space in \cite{oystein}. Combining CMB and SNIa, in
\cite{oystein} they
got tighter constraint on $\alpha$, especially the
lower bound on the
negative value of $\alpha$, which calls for further
understanding in our
study. Comparing with the investigation of the
observational constraint,
we found that holography again plays a powerful role
in the study of
gravity.

ACKNOWLEDGEMENT: This work was partically supported
by  NNSF, China,
Ministry of Science and 
Technology of China under Grant No. NKBRSFG19990754
, Ministry of
Education of China and Shanghai Education Commission. The work of E.A. was supported by 
Fundac\~ao de Amparo \`a Pesquisa do Estado de
S\~ao Paulo (FAPESP) and Conselho Nacional de
Desenvolvimento
Cient\'{\i}fico e Tecnol\'ogico (CNPQ).


\begin{thebibliography}{x}
\bibitem{1} J. D. Bekenstein, {\it Phys. Rev.} {\bf
 D 23}, 2817(1981).
\bibitem{2} W. G. Unruh and R. M. Wald, {\it Phys.
Rev. }{\bf D 25}, 942
(1982); ibid {\bf 27}, 2271 (1983). 
\bibitem{3} J. D. Bekenstein, {\it Phys. Rev.}{\bf
D49}, 1912 (1994); ibid
{\bf 60}, 124010 (1999); M. Schiffer and J. D.
Bekenstein, {\it
Phys. Rev.}{\bf D 39}, 1109 (1989); O. B.
Zaslavskii, {\it
Class. Quan. Grav.} {\bf 13}, L7 (1996); J. D.
Bekenstein and A. E. Mayo,
{\it Phys. Rev.}{\bf D 61}, 024022 (2000); S. Hod,
{\it Phys. Rev. }{\bf D
61}, 024023 (2000); B. Linet, {\it Gen Relativ.
Grav.} {\bf 31}, 1609
(1999); B. Linet, {\it Phys. Rev. }{\bf D 61},
107502 (2000); B. Wang and
E. Abdalla, {\it Phys. Rev.}{\bf D 62}, 044030
(2000); W. Qiu, B. Wang,
R. K. Su and E. Abdalla, {\it Phys. Rev. }{\bf D
64}, 027503 (2001). 
\bibitem{4} J. D. Bekenstein, {\it Inter. J. Theor.
Phys} {\bf 28}, 967
(1989). 
\bibitem{5} G. 't Hooft, gr-qc/9310026; L. Susskind,
{\it
  J. Math. Phys. } {\bf 36} (1995) 6377.
\bibitem{6} W. Fischler and L. Susskind, hep-th/9806039.
\bibitem{7} N. Kaloper and A. Linde, {\it  Phys.
Rev. } {\bf D 60 } (1999)
103509;  R. Easther and D. A. Lowe, {\it  Phys. Rev.
Lett. } {\bf 82 }
(1999) 4967;  R. Brustein, {\it  Phys. Rev. Lett. }
{\bf 84} (2000) 2072;
R. Brustein, G. Veneziano, {\it  Phys. Rev. Lett. }
{\bf 84} (2000) 5695;
R. Bousso,  {\it JHEP } {\bf 7} (1999) 4, ibid {\bf
6} (1999) 28,  {\it
Class. Quan. Grav.} {\bf  17} (2000) 997; B. Wang,
E. Abdalla, {\it
Phys. Lett. } {\bf B 466} (1999) 122,  {\bf B 471}
(2000) 346; B. Wang,
E. Abdalla and R. K. Su, {\it Phys. Lett.}{\bf B
503}, 394 (2001). 
\bibitem{8} G. Veneziano, {\it  Phys. Lett. } {\bf
B454}(1999) 22;
G. Veneziano, hep-th/9907012; E. Verlinde,
hep-th/0008140. 
\bibitem{9} R. Tavakol, G. Ellis, {\it  Phys. Lett.
} {\bf B 469} (1999) 37.
\bibitem{10} B. Wang, E. Abdalla and T. Osada, {\it
 Phys. Rev. Lett. }
{\bf 85} (2000) 5507. 
\bibitem{11} I. Savonije and E. Verlinde, {\it Phys.
Lett.}{\bf B 507}
  (2001) 305; Bin Wang, Elcio Abdalla and Ru-Keng Su
{\it
  Mod. Phys. Lett.} {\bf A17} (2002) 23; S. Nojiri
and S. D. Odintsov,
  Int. J. Mod. Phys A16 (2001) 3237;  D. Kutasov, F.
Larsen, JHEP 0101
  (2001) 001, hep-th/0009244; F. Lin, Phys. Lett. B
507 (2001) 270;
  R. Brustein, S. Foffa and G. Veneziano, Nucl.
Phys. B 601 (2001) 380;
  D. Klemm, A. C. Petkou and G. Siopsis,
hep-th/0101076; R. G. Cai, 
  Phys. Rev. D 63 (2001) 124018 hep-th/0102113; D.
Birmingham and
  S. Mokhtari, Phys. Lett. B 508 (2001) 365
hep-th/0103108.  
\bibitem{12} E. Witten, {\it Science} {\bf 285}, 512
(1999). 
\bibitem{13} P. Horava and D. Minic, {\it Phys. Rev.
Lett. }{\bf 85}, 1610
(2000). 
\bibitem{14} T. Banks and W. Fischler
astro-ph/0307459; B. Wang and
E. Abdalla, {\it Phys. Rev.}{\bf D} (in press),
hep-th/0308145; R. G. Cai,
JCAP 0402:007,2004; D. A. Lowe and D. Marolf,
hep-th/0402162. 
\bibitem{15} S. Perlmutter et al, {\it Astrophys.
J}{\bf 517}, 565 (1999);
A. Riess et al, {\it Astro J}{\bf 116}, 1009 (1998). 
\bibitem{16} A. Cohen, D. Kaplan and A. Nelson, {\it
Phys. Rev. Lett. }{\bf 82}, 4971 (1999). 
\bibitem{17} S. D. H. Hsu, hep-th/0403052.
\bibitem{18} M. Li, hep-th/0403127.
\bibitem{19} G. Dvali, G. Gabadadze and M. Porrati,
{\it Phys. Lett.}{\bf
B 485}, 208 (2000); C. Deffayet, {\it Phys.
Lett.}{\bf B 502}, 199 (2001);
C. Deffayet, G. Dvali and G. Gabadadze, {\it Phys.
Rev.}{\bf D 65}, 044023
(2002). 
\bibitem{20} G. Dvali and M. S. Turner,
astro-ph/0301510.
\bibitem{21} E. V. Linder, astro-ph/0402503.  
\bibitem{oystein} O. Elgaroy and T. Multamaki,
astro-ph/0404402. 
\bibitem{23} S. M. Carroll, W. H. Press and E. L.
Turner, {\it
 Ann. Rev. Astron. Astrophys.} {\bf 30} (1992) 499-542;
 R.R. Caldwell, R. Dave, P. J. Steinhardt {\it Phys.
Rev. Lett.} {\bf 80}
 (1998)) 1582-1585, astro-ph/9708069.
\bibitem{ray} D. Bak, S. J. Rey, Class.Quant.Grav.
17 (2000) L83

\end{thebibliography}
\end{document}